\documentclass[epsfig,12pt]{article}

\newcommand {\eqref} [1] {(\ref {#1})}
\newcommand {\slsh} [1] {\not{\hbox{\kern-2pt${#1}$}}}

\newcommand {\beq} {\begin{equation}}
\newcommand {\eeq} {\end{equation}}
  \newcommand {\ber}{\begin{eqnarray*}}
  \newcommand {\eer} {\end{eqnarray*}}
\newcommand {\bea}{\begin{eqnarray}}
  \newcommand {\eea} {\end{eqnarray}}

\newcommand{\ntwo}{${\cal N}=2\ $}

\newcommand{\Dslash}{\,{\raise.15ex\hbox{/}\mkern-12mu D}}

\newcommand{\gsim}{\lower.7ex\hbox{$
\;\stackrel{\textstyle>}{\sim}\;$}}
\newcommand{\lsim}{\lower.7ex\hbox{$
\;\stackrel{\textstyle<}{\sim}\;$}}
\newcommand{\pt}{\partial}
\newcommand{\vp}{\varphi}

\def\beqn{\begin{eqnarray}}
\def\eeqn{\end{eqnarray}}

\begin{document}
\begin{titlepage}
\begin{flushright}{UMN-TH-2627/07\,, FTPI-MINN-07/36\,, }
\end{flushright}

\vskip 0.5cm

\centerline{{\Large \bf Non-Abelian Strings and the L\"uscher Term}}

\vskip 1cm
\centerline{\large M. Shifman,${}^{a}$ and A. Yung${}^{a,b,c}$ }

\vskip 0.5cm

\begin{center}

\vspace{5mm}

$^a${\it  William I. Fine Theoretical Physics Institute,
University of Minnesota,
Minneapolis, MN 55455, USA}\\[1mm]
$^{b}${\it Petersburg Nuclear Physics Institute, Gatchina, St. Petersburg
188300, Russia\\[1mm]
$^c${\it Institute of Theoretical and Experimental Physics, Moscow
117259, Russia}}
\end{center}

\vskip 1cm

\begin{abstract}

We calculate the L\"uscher term for recently suggested non-Abelian flux tubes (strings).
The main feature of the non-Abelian strings is the presence of orientational
zero modes associated with rotation of their color flux inside a
non-Abelian subgroup. The L\"uscher term is determined by the number of light degrees
of freedom  on the string wordsheet. Unlike the standard $\pi/12$ we get
$N\pi/12$ for non-Abelian strings in the U$(N)$ gauge theories.
Thus, the L\"uscher coefficient acquires a dependence on the rank of the gauge group.
In the models with non-Abelian strings 
discussed in the literature there are two distinct scales:
the string tension $\xi$ (the string thickness $\sim \xi^{-1/2}$)
and the dynamical scale of strong interactions
$\Lambda$. At weak coupling $\xi\gg\Lambda^2$.
The  L\"uscher term for non-Abelian strings experiences a jump:
at $\xi^{-1/2}\ll L\ll \Lambda^{-1}$ it is $N\pi/12$ while at
at $L\gg \Lambda^{-1}$
the orientational moduli are frozen out and 
the L\"uscher coefficient approaches its ``L\"uscher" value
$\pi/12$. We raise the question of possible extra (i.e. non-translational)
light moduli 
on the worldsheet of QCD strings at large $N$.

\end{abstract}

\end{titlepage}

\section{Introduction}

The energy of a long string (flux tube) in confining gauge theories
behaves as
\beq
E(L) = T\,L + C -\frac{\gamma}{L} + ...
\label{en}
\eeq
where $L$ is the string length, $C$ is a constant of dimension of mass while
$\gamma$ is a dimensioneless constant. The $O(1/L)$ term
is referred to as the L\"uscher term \cite{lus}.
Its value was calculated by L\"uscher,
\beq
\gamma= \frac{\pi}{12}\,,
\label{gam}
\eeq
and is believed to be universal. In fact, the L\"uscher coefficient measures the number
of light (massless) degrees of freedom on the string world sheet.
Equation (\ref{gam}) assumes that the only massless excitations of the string
are due to two translational zero modes. 

Recently discovered non-Abelian strings \cite{nus} do not satisfy this assumption.
What is the difference between Abelian \cite{ANO} and non-Abelian strings?
In the former case the gauge group acting in the infrared and responsible for the flux tube formation is Abelian (i.e. U(1)$\times$U(1) ...). In the latter case
we deal with a non-Abelian group in the infrared.
In addition to the position of the string center in the perpendicular plane,
non-Abelian strings are characterized by internal moduli.
 The best-known example of the first type is the Seiberg--Witten string found in \cite{Seiberg:1994rs} in a slightly deformed
${\mathcal N}=2$ super-Yang--Mills theory. If the deformation parameter $\mu$ is small,
$$
\mu\ll \Lambda \,,
$$
the SU$(N)$ gauge group is spontaneously broken, the group acting in the low-energy description
is U(1)$^{N-1}$, and the string obtained is a generalization of 
the good old Abrikosov flux tube \cite{ANO}. It is Abelian.

In the opposite limit
$$
\mu\gg\Lambda \,,
$$
the breaking of SU$(N)$ down to U(1)$^{N-1}$ does not occur.
The infrared dynamics is determined by SU$(N)$; the corresponding flux tube should
be 
non-Abelian. Presumably, there is no phase transition in $\mu$,
and the Abelian and non-Abelian flux tubes are smoothly connected.
A similar phenomenon takes place 
\cite{AMM}  in QCD-like theories on $R_3\times S_1$. The radius $r$
of the compactified dimension plays the same role as $\mu$.
Unfortunately, the limit $\mu\gg\Lambda$  or $r\gg \Lambda^{-1}$ are not under 
theoretical control. The first {\em non-supersymmetric} example of a controllable 
situation,
in which a non-Abelian string emerges at weak coupling,
was discussed in \cite{GSY}. In this model there
are two distinct scales:
the string tension $\xi$ (the string thickness $\sim \xi^{-1/2}$)
and the dynamical scale of strong interactions
$\Lambda$. At weak coupling $\xi\gg\Lambda^2$. 
The L\"uscher coefficient measures the number of light degrees of freedom
on the string worldsheet.
The main feature of the non-Abelian strings is the presence of orientational
zero modes associated with rotation of their color flux inside a
non-Abelian subgroup.  Unlike the standard $\pi/12$ we get
$N\pi/12$ for the non-Abelian strings in the U$(N)$ gauge theories.
Thus, the L\"uscher coefficient acquires a dependence on the rank of the gauge group.
In fact, 
the  L\"uscher term for non-Abelian strings experiences a jump:
at $\xi^{-1/2}\ll L\ll \Lambda^{-1}$ it is $N\pi/12$ while at
at $L\gg \Lambda^{-1}$
the orientational moduli are frozen out and 
the L\"uscher coefficient approaches its ``L\"uscher" value
$\pi/12$. 

At the end of this paper we discuss the question of possible extra (i.e. non-translational)
light moduli 
on the worldsheet of QCD strings at large $N$.

\section{A model supporting non-Abelian strings}

The model discussed in \cite{GSY} is a U$(N)$ gauge theory with
$N$ flavors of complex scalars (squarks) $\vp^A$ ($A=1,...,N$) in the fundamental 
representation of the gauge group.
The action of
this model is
\beqn
S &=& \int {\rm d}^4x\left\{\frac1{4g_2^2}
\left(F^{a}_{\mu\nu}\right)^{2}
+ \frac1{4g_1^2}\left(F_{\mu\nu}\right)^{2}
+ |\nabla^\mu \vp^A |^2
 \right.
 \nonumber\\[3mm]
&+&
\left.
 \frac{g^2_2}{2}
\left(\bar{\vp}_A T^a \vp^A \right)^2
 +
 \frac{g^2_1}{8}\left(|\vp^A|^2- N\xi \right)^2
\right\}\,,
\label{nsusymodel}
\eeqn
where 
$$
\nabla_{\mu}=\pt_{\mu} -\frac{i}{2}\,A_{\mu} -iT^a\,A_{\mu}^a\,,
$$
$A_{\mu}$ and 
$A^a_{\mu}$ are the U(1) and SU$(N)$ gauge fields, respectively,
 $T^a$ are the SU$(N)$ generators, while  $g_1^2$ and $g_2^2$ are 
 the U(1) and SU$(N)$ gauge 
couplings.

Squark fields develop vacuum expectation values (VEV's) of 
 the color-flavor locked form
\beqn
\langle q^{kA}\rangle &=&\sqrt{\xi}\,
\left(
\begin{array}{ccc}
1 & \ldots & 0 \\
\ldots & \ldots & \ldots\\
0 & \ldots & 1\\
\end{array}
\right),
\nonumber\\[4mm]
k&=&1,..., N\qquad A=1,...,N\, ,
\label{qvev}
\eeqn
where the squark fields are written  as an $N\times N$ matrix in 
the color-flavor indices.
The  VEV's (\ref{qvev})  spontaneously
break both the gauge and flavor SU($N$)'s.
A diagonal global SU($N$) survives, however,
\beq
{\rm U}(N)_{\rm gauge}\times {\rm SU}(N)_{\rm flavor}
\to {\rm SU}(N)_{C+F}\,.
\label{c+f}
\eeq
Thus, a color-flavor locking takes place in the vacuum.
A version of this pattern of the symmetry breaking was suggested
long ago \cite{BarH}.
Since the gauge symmetry is broken, both the U(1) and SU($N$) gauge bosons acquire
masses
\beq
m_{{\rm U}(1)}=g_1\sqrt{\xi},\;\;\;\;\;\; m_{{\rm SU}(N)}=g_2\sqrt{\xi}\,,
\label{gmasses}
\eeq
respectively.

This model supports string solutions \cite{GSY}, which break
global SU($N$)$_{C+F}$ symmetry present in the vacuum down to 
SU($N-1)\times$ U(1). This ensures appearance of orientational zero modes
of the string  making it non-Abelian. The phenomenon is quite
similar to supersymmetric 
non-Abelian strings found in \ntwo supersymmetric gauge theories \cite{nus}.
The tension of the elementary string is determined by the squark VEV's \cite{GSY},
\beq
T_{\rm cl}=2\pi\xi\,,
\label{ten}
\eeq
where the subscript cl marks the classical approximation.
The orientational moduli belong to the quotient
\beq
CP(N-1)=\frac{{\rm SU}(N)_{C+F}}{{\rm SU}(N-1)\times {\rm U}(1)}\,.
\label{moduli}
\eeq
The low-energy effective theory on the string worldsheet 
is given by the $CP(N-1)$ model with the action 
\beq
S^{(1+1)}=   \int d^2 x \,\left\{\frac{T_{\rm cl}}{2}(\pt_k z^i)^2+
2 \beta\,\left|\nabla_k n^l\right|^2
\right\}\,,
\label{cpN}
\eeq
(see \cite{GSY} or the  review  paper \cite{SYrev} for derivation).
 Here $k=1,2$ labels the worldsheet coordinates,
$z^i$, $i=1,2$ are two real translational moduli (the string position 
in the plane orthogonal to the string) and $N$ complex fields $n^l$ ($l=1,...,N$)
subject to the constraint  $|n^l|^2=1$ denote orientational moduli. The axillary U(1)
gauge field gauging the 
common U(1) phase of $n^l$ enters without kinetic energy, and  $\nabla_k=\pt_k-iA_k$. The two-dimensional coupling $\beta$ is related to
the bulk coupling, $\beta=2\pi/g^2_2$ at the scale $\sqrt{\xi}$.
Overall we have $2N-1-1=2(N-1)$ real orientational
moduli --- we subtracted one constraint and one ``eaten'' phase. This is 
the number of degrees of freedom in the  $CP(N-1)$ model.

This example taught us that, besides two translational gapless excitations,
other light modes associated with internal degrees of freedom, can exist on the string worldsheet.The  $CP(N-1)$ model part of the worldsheet theory
 gives rise
to the scale parameter $\Lambda$ which determines the mass gap
for orientational moduli.
The tension of the string $\xi$ is much larger than $\Lambda^2$.
This implies that there exists a window of distances,
\beq
\xi^{-1/2} \ll L \ll \Lambda^{-1}
\label{win}
\eeq
in which all $2N$ moduli characterizing the non-Abelian string 
--- two translational and $2N-2$ orientational ---
can be
considered as massless. Correspondingly, in this window
the L\"uscher coefficient will take the value
\beq
\gamma = \frac{N\pi}{12}\,.
\label{nlu}
\eeq
In the transitional domain $L\sim \Lambda^{-1}$ it must smoothly decrease
eventually tending to (\ref{gam}) at $L\gg \Lambda^{-1}$.

\section{The L\"uscher coefficient}

To derive (\ref{nlu}) we calculate the quantum energy of the string
of the length $L$ with nailed ends. We have
\beq
E_{\rm qu}=\sum_k N\,\frac{\pi k}{L}\,,
\label{energy}
\eeq
where the subscript qu indicates the inclusion of quantum corrections
(we are interested only in the infrared corrections).
The above expression is divergent. To regularize it we introduce a ``lattice''
spacing $\epsilon$ ($1/\epsilon$ is the ultraviolet (UV) cutoff). Then we have
\beqn
E_{\rm qu} &=&\sum_k N\frac{\pi k}{L}\exp{\left(-\frac{\pi k}{L}\,\epsilon\right)}
\label{renergy}
\\[3mm]
 &=& \frac{N}{\pi\epsilon^2}\,L -  \frac{N\pi}{12}\,\frac1L + O(\epsilon).
\label{quenergy}
\eeqn

What is the meaning of the first divergent term here? It is 
easy to see that it presents renormalization
of the classical string tension (\ref{ten}). To phrase it
differently,  the $O(1/\epsilon^2)$ term  is the one-loop contribution
in the  vacuum energy in the worldsheet  theory (\ref{cpN}).
The $CP(N-1)$ model (\ref{cpN}) is an  effective theory which describes
dynamics of the string light modes at  energies below the inverse thickness
of the string given by the bulk theory masses (\ref{gmasses}). 
Thus the UV cutoff $1/\epsilon$ in this theory must be  of the order of $g\sqrt{\xi}$,
where we assume that both coupling constants in (\ref{nsusymodel}) are of the 
same order, $g^2_1\sim g^2_2\sim g^2$. Thus, we conclude
that  the total string energy is
\beq
E= \left(2\pi\xi+ {C}\,g^2\xi\right)\,L -  
\frac{N\pi}{12}\,\frac1L + O(1/L^2).
\label{totenergy}
\eeq 

We can use this result at  $L$ much larger than the string thickness,  $L\gg 1/(g\sqrt{\xi})$. It has a different $L$ dependence
 as compared to the leading linear in $L$ term. Therefore, it makes
sense to trace this term. At $L\gg \Lambda^{-1}$ 
the orientational degrees of freedom freeze out and no longer contribute to the string energy.

In fact, the  L\"uscher coefficient in non-Abelian strings
can follow even a richer pattern of behavior. Indeed,
the models \cite{nus} admit another dimensional parameter in the bulk
$\Delta m$. This parameter manifests itself on the world sheet
as a twisted mass. Instead of $CP(N-1)$ model for orientational moduli,
we get $CP(N-1)$ with the twisted mass. If
\beq
\xi \gg |\Delta m| \gg \Lambda
\eeq
the orientational modes acquire the mass gap 
equal to $ |\Delta m| $ (see \cite{nus,GSY}).
This implies, in turn, that the window (\ref{win}) is divided into two sub-windows,
$\left[\xi^{-1/2} \,,\,\,|\Delta m|^{-1}\right]$ and
$\left[|\Delta m|^{-1}\,,\,\, \Lambda^{-1}\right]$.
In the first sub-window the L\"uscher coefficient is given by (\ref{nlu})
while in the second by (\ref{gam}). 
This parameter can be adjusted at will.
As $|\Delta m|\to 0$ the second sub-window shrinks to nothing.

\section{QCD strings}

The most intriguing question is whether or not
the QCD string and strings in other QCD-like theories has 
only two translational massless moduli on its worldsheet.
Of course, at $N=3$ the emergence of other light modes
(with the mass mass gap $\ll \Lambda$) does not seem likely
since the only dimensional parameter in this case is $\Lambda$.
Moreover, lattice numerical data (see e.g. Greensite's review
quoted in \cite{lus}) support the L\"uscher value (\ref{gam}).
However, at $N\gg 1$ the answer does not seem so obvious.
A priori it is not ruled out that in the multicolor limit
the QCD string acquires an analog of the orientational moduli
of Refs. \cite{nus,GSY} with the mass gap suppressed by powers of $1/N$.
A heuristic motivation is provided by a consideration 
on $R_3\times S_1$ similar to \cite{AMM}. If the value of the
$S_1$ radius $r$ is small
compared to $\Lambda^{-1}$, we deal with $N-1$ distinct Abelian strings
which must fuse into a single non-Abelian string at 
$r> \Lambda^{-1}$. Additional (quasi) moduli might occur
in the process of fusion.

Light modes are definitely present on the worldsheet of $k$-strings
\cite{AS}. 
The issue of existence/nonexistence
of ``light" modes of the fundamental QCD strings,
unrelated to the excitations of the internal flux-tube structure,
is hard to investigate in a model-independent way since
the underlying dynamics is that of strong coupling.
The answer could be provided by precision lattice measurements
in the multicolor Yang--Mills. The question of feasibility of such measurements
remains open.

\section{Discussion}

A standard presumption in string theory is that
the theory on the string worldsheet must be conformal.
This assumption is also applied to
long-distance dynamics of string solitons in four-dimensional field
theories, see e.g. \cite{PS}. Non-Abelian strings discussed in this paper
present a clear-cut counterexample.
The low-energy dynamics on the string worldsheet
is non-conformal because so is the $CP(N-1)$ model.
The total number of the moduli fields is $2N$ rather than two
translational moduli. This implies that the standard L\"uscher coefficient
$\pi/12$ changes to $N\pi/12$ in the window (\ref{win}).
An open and intriguing question is whether QCD strings 
at large $N$ acquire light (quasi)moduli additional with regards to
two
translational moduli.

\vspace{0.5cm}

We are grateful to J. Maldacena, N. Seiberg, E. Witten and
other participants of the HEP theory seminar at Princeton University for stimulating questions, and to D. Tong for a remark.
This work  is supported in part by DOE grant DE-FG02-94ER408. 
The work of A.Y. was  supported 
by  FTPI, University of Minnesota, 
by RFBR Grant No. 06-02-16364a 
and by Russian State Grant for 
Scientific Schools RSGSS-11242003.2.

\end{document}